\documentclass[a4,12pt]{article}
\usepackage{fullpage}
\usepackage{enumerate}
\usepackage{amsmath}
\usepackage{amssymb}
\usepackage{amsthm,amstext}
\usepackage{graphicx}
\usepackage{mathtools}
\usepackage{units}
\usepackage{comment}

\title{Bridging kinetic plasma descriptions and single fluid models}
\author{A. Crestetto$^{\star}$, F. Deluzet$^{\dagger}$, D. Doyen$^\ddagger$ \\[1em]
$^{\star}$Laboratoire de Math\'ematiques Jean Leray UMR 6629,\\
2, rue de la Houssini\'ere,\\
 F-44322 Nantes Cedex 3, France,\\
anais.crestetto@univ-nantes.fr\\[1em]
$^\dagger$Universit\'e de Toulouse; UPS, INSA, UT1, UTM,\\
 Institut de Math\'ematiques de Toulouse,\\
 CNRS, Institut de Math\'ematiques de Toulouse UMR 5219,\\
F-31062 Toulouse, France,\\
fabrice.deluzet@math.univ-toulouse.fr \\[1em]
$^\ddagger$Universit\'e de Marne-la-Vall\'ee,\\ Laboratoire d'Analyse et de Math\'ematiques Appliqu\'ees,\\
CNRS, Laboratoire d'Analyse et de Math\'ematiques Appliqu\'ees UMR 8050,\\
5, boulevard Descartes, Cit\'e Descartes - Champs-sur-Marne,\\
F-77454 Marne-la-Vallée, France,\\
david.doyen@univ-mlv.fr
}

\def\vHall{\text{v}_\text{H}}
\def\Me{\mathcal{M}_e}
\def\PiMe{\Pi_{\Me}}
\def\Id{\mathbb{I}\textnormal{d}}
\def\LMe{\mathcal{L}_{M_e}}

\begin{document}

\maketitle

\begin{abstract}
  The purpose of this paper is to bridge kinetic plasma descriptions and low frequency single fluid models. More specifically, the asymptotics leading to Magneto-Hydro-Dynamic (MHD)  regimes starting from the Vlasov-Maxwell system are investigated. The analogy with the derivation, from the Vlasov-Poisson system, of a fluid representation for the ions coupled to the Boltzmann relation for electrons is also outlined. The aim is to identify asymptotic parameters explaining the transitions from one microscopic description to a macroscopic low frequency model. These investigations provide ground work for the derivation of multi-scale numerical methods, model coupling or physics based preconditioning.
\end{abstract}

\section{Introduction}
The aim of this paper is to propose a continuation of the work initiated in \cite{DDN06,DDNSV10,DegDelDoy} focusing on the derivation of asymptotic preserving schemes for kinetic plasma descriptions in the quasi-neutral limit. The purpose of these numerical methods is to provide a quasi-neutral description of the plasma with no constraints on the simulation parameters related to the Debye length but with the ability to perform local up-scalings with non neutral plasma descriptions. This brings a gain in the computational efficiency, since the discretization parameters can be set according to the physics of interest rather than the small scales (namely the Debye length) described by the model.

The methodology introduced in these former achievements is aimed to be generalized here to more singular limits. In this series of former works, the limit models remain kinetic and the scales of interest are related to the electron dynamics. For instance, the quasi-neutral limit of the Vlasov-Maxwell system investigated in \cite{DegDelDoy} can be interpreted as a kinetic description of the Electron-Magneto-Hydro-Dynamic (E-MHD) \cite{gordeev_electron_1994,swanekamp_particlecell_1996,cho_anisotropy_2004}, accounting for the electron inertia, the massive ions being assumed at rest or slowly evolving. In the present paper, the objective is to go beyond the kinetic E-MHD with the aim to bridge the Vlasov-Maxwell system and Magneto-Hydro-Dynamic (MHD) models. In MHD systems, the scales of interest are defined by the overall plasma dynamic which is governed by the ions, the fast scales associated to the electron inertia being filtered out from the equations. 

 This preliminary work is therefore devoted to the derivation of a model hierarchy bridging either the Vlasov-Maxwell system and MHD models for magnetized plasmas or, the Vlasov-Poisson and the electron adiabatic response also referred to as Boltzmann relation (see \cite{langmuir_interaction_1929,tonks_oscillations_1929,tonks_general_1929} for seminal works and \cite{de_cecco_asymptotic_2017} for numerical investigations), for electrostatic frameworks. A wide range of applications of the present investigations can be named, specifically low variance Particle-In-Cell methods or more generally numerical discretization of kinetic models implementing a Micro-Macro decomposition of the distribution function. We refer for instance to \cite{crestetto_kinetic/fluid_2012,crouseilles_asymptotic_2011,dimarco_asymptotic_2014,lemou_new_2008}) for Micro-Macro methods, and to \cite{degond_moment-guided_2011} for the moment Guided method; fluid-preconditioned fully implicit methods \cite{chen_energy-_2011,chen_fluid_2014,chen_multi-dimensional_2015} and Asymptotic-Preserving numerical methods \cite{Jin99,degond_asymptotic-preserving_2017}. Another application can be envisioned with the hybrid coupling of Particle-In-Cell methods and MHD descriptions \cite{schumer_mhd--pic_2001,daldorff_two-way_2014} and more generally coupling strategies such as the Current-Coupling-Scheme (CCS) and the Pressure-Coupling-Scheme (PCS) (see \cite{park_threedimensional_1992} \cite{tronci_hybrid_2014} and the references therein).

 The aim of this work is to clarify how the asymptotic parameters interact with each other and define reduced models, but also, to relate these parameters to meaningful physical quantities. The MHD regime is sometimes derived by letting $\varepsilon_0$, the vacuum permittivity, go to zero (see for instance \cite{jang_derivation_2012,tronci_neutral_2015}) which is referred to as the full Maxwell to the low frequency pre-Maxwell's equations asymptotic in \cite[see section~2.3.3]{freidberg_ideal_2014}. It is also common to let the electron to ion mass ratio go to zero to explain the vanishing of the electron inertia (\cite{freidberg_ideal_2014,klingenberg_consistent_2016}) in deriving either MHD modelling or the Boltzmann relation. Although the right asymptotic models are recovered by this means, these assumptions do not account for changes in the system characteristics that may explain for a regime transition: the electron to ion mass ratio remains constant and the same property holds true for the vacuum permittivity.

The outlines of the paper are the following.
The plasma kinetic description is introduced in Sec.~\ref{sec:eq:cont} together with the Maxwell system driving the evolution of the electromagnetic field. A dimensionless form of the system is stated  in order to develop an asymptotic analysis and the derivation of reduced models. A hierarchy of quasi-neutral model is proposed in Sec.~\ref{sec:fluid:hierarchy} for the Vlasov-Maxwell system. It encompasses fully kinetic, hybrid as well as single fluid (MHD) plasma descriptions. The electrostatic framework is investigated in Sec.~\ref{sec:electrostatic}. The electrostatic limit of the Maxwell system is performed. A hierarchy of models, similar to that of the electromagnetic framework is derived. Finally, a synthesis of these asymptotic analysis is proposed in Sec.~\ref{sec:conclusions} devoted to conclusions.

\section{The Vlasov-Maxwell system in a dimensionless form}\label{sec:eq:cont}
\subsection{Introduction}
In this section, the purpose is to unravel a series of asymptotic limits bridging gap between the Vlasov-Maxwell system  and a Magneto-Hydro-Dynamic (MHD) model. The difficulty is therefore to identify parameters explaining the transition from one description to the other and to relate these parameters to specific characteristics of the system. The tools mobilize to meet this aim are based on the asymptotic analysis of the Vlasov-Maxwell system. Since the low frequency plasma modelling is related to a fluid plasma description, the kinetic model is upgraded with collision operators. Therefore, the most refined modelling consists of a Valsov equation for the electrons and the ions, augmented with a collision operator and coupled to the Maxwell system. Even if the physical model is non collisional or weakly collisional, the transition towards a fluid limit is accounted for by a collisional process, thanks to a BGK operator. This choice of collision operator is questionable from a strict modelling view point, nonetheless, the purpose here is to easily derive the fluid limit at a limited computational cost. In this respect the BGK collision operator is a good candidate. 
First, the whole collisional processes are considered, including both inner and inter species collisions. Nonetheless, only the minimal collisional process will be accounted for to derive a MHD regime from the kinetic model. 
This point will be outlined in the following sections. The introduction of non dimensional quantities will naturally reveal dimensionless parameters in the equations. Letting some of these parameters go to zero shapes the hierarchy of models derived for the Vlasov-Maxwell system and bridging the gap with MHD models.
\subsection{The Vlasov-BGK-Maxwell system}
\label{sec:model}
The most refined description of the plasma is constituted by two Vlasov equations, $f_i$ and $f_e$ being the ion and electron distribution functions
\begin{align}
  & \partial_t f_i +v \cdot \nabla_x f_i  + \frac{q}{m_i} \left( E + v \times B\right) \cdot \nabla_v f_i =  \mathcal{Q}_i \,, \\
 & \partial_t f_e +v \cdot \nabla_x f_e - \frac{q}{m_e} \left( E + v \times B \right)\cdot  \nabla_v f_e =  \mathcal{Q}_e \,.
\end{align}
In these equations, $q$ is the elementary charge, $m_\alpha$ is the mass of the species $\alpha$ ($\alpha=e$ for the electrons and $i$ for the ions). The BGK  collision operator $\mathcal{Q}_\alpha$ are given by \cite{huba_nrl_2011}
\begin{equation}
  \begin{split}
  \mathcal{Q}_\alpha &=  \mathcal{Q}_{\alpha\alpha}+ \mathcal{Q}_{\alpha\beta} \,,\\
  \mathcal{Q}_{\alpha\alpha} = {\nu_{\alpha\alpha}} \left( \mathcal{M}_{n_\alpha,u_\alpha,T_\alpha} - f_\alpha \right)\,, &\qquad 
   \mathcal{Q}_{\alpha\beta}={\nu_{\alpha\beta}} \left( \overline{\mathcal{M}}_{n_\alpha,\overline{u}_\beta,\overline{T}_\beta} - f_\alpha \right) \,,    
  \end{split}
\end{equation}
$\nu_{\alpha\alpha}$ and $\nu_{\alpha\beta}$ being the like-particle and inter species collision frequencies which can be defined as \cite{degond_chapter_2007,spatschek_high_2012}
\begin{subequations}\label{eq:Freq:Ordering}
\begin{alignat}{3}
  \nu_{ii} &= K_0 \, \frac{n_i}{(k_B T_i)^\frac{3}{2}}\frac{\sqrt{2}}{\sqrt{m_i}}\,, \qquad \nu_{ie} &&= K_0 \, \frac{n_e}{(k_B T_i)^\frac{3}{2}}\frac{\sqrt{m_e}}{m_i}\,, \\
  \nu_{ee} &=  K_0 \, \frac{n_e}{(k_B T_e)^\frac{3}{2}}\frac{\sqrt{2}}{\sqrt{m_e}}\,, \qquad \nu_{ei}&&= K_0 \, \frac{n_i}{(k_B T_e)^\frac{3}{2}}\frac{1}{\sqrt{m_e}} \,, 
\end{alignat}  
where, $C$ denotes a constant with a magnitude equal to one, $\ln(\Lambda)$ the Coulomb logarithm and the ions being assumed mono-charged, 
\begin{equation}
  K_0 =  C \left(\frac{q^2}{4 \pi \epsilon_0}\right)^2 \ln(\Lambda) \,.
\end{equation}
\end{subequations}

The Maxwellians $\mathcal{M}_{n_\alpha,u_\alpha,T_\alpha}$ and $\overline{\mathcal{M}}_{n_\alpha,\overline{u}_\beta,\overline{T}_\beta}$ are defined as
\begin{subequations}\label{eq:def:BGK}
\begin{align}
  \mathcal{M}_{n_\alpha,u_\alpha,T_\alpha}  &= n_\alpha(x,t) \left(\frac{ m_\alpha}{2\pi k_B T_\alpha(x,t)}\right)^\frac{D_v}{2} \exp\left(-\frac{m_\alpha|u_\alpha(x,t)-v |^2}{2 k_B T_\alpha(x,t)} \right) \,,\\
  \overline{\mathcal{M}}_{n_\alpha,\overline{u}_\beta,\overline{T}_\beta}  &= n_\alpha(x,t) \left(\frac{ m_\alpha}{2\pi k_B \overline{T}_\beta(x,t)}\right)^\frac{D_v}{2} \exp\left(-\frac{m_\alpha|\overline{u}_\beta(x,t)-v |^2}{2 k_B \overline{T}_\beta(x,t)} \right) \,, \label{eq:BGK:inter}
\end{align}
\end{subequations}
 $D_v$ denoting the dimension of the velocity space, $k_B$ is the Boltzmann constant. The Maxwellian parameters are $n_\alpha$, $u_\alpha$ and $T_\alpha$ the density, mean velocity and temperature associated to the distribution function $f_\alpha$ and defined as
\begin{equation}
  n_\alpha = \int_{\Omega_v} f_\alpha dv \,, \qquad n_\alpha u_\alpha= \int_{\Omega_v} v \, f_\alpha \, dv \,, \qquad \frac{1}{\gamma-1}n_\alpha k_B T_\alpha = \frac{m_\alpha}{2}\int_{\Omega_v} |v-u|^2\, f_\alpha  \, dv \,,
\end{equation}
with $\gamma$ the specific heat ratio whose value depends on the dimensionality of the velocity space $D_v$ through
\begin{equation}
  \gamma - 1 = \frac{2}{D_v} \,.
\end{equation}
The collision operators verify the following conservation properties 
\begin{subequations}
  \begin{alignat}{3}
    &\int \mathcal{Q}_{\alpha\alpha} \, m_\alpha v^n \, dv = 0  \,, \qquad &&n = 0,\ldots,2\,, \\
    &\int \mathcal{Q}_{\alpha\beta} \, m_\alpha v^n \, dv   + \int \mathcal{Q}_{\beta\alpha}\,  m_\beta v^n\,   dv = 0 \,, \qquad &&n = 0,\ldots,2 \,. \label{eq:prop:BGK:inter}
  \end{alignat}
\end{subequations}

The temperature and the mean velocity $(\overline{u}_\beta,\overline{T}_\beta)$ in the inter-specie collision operator expression \eqref{eq:BGK:inter} should be chosen with care in order to guarantee the total momentum and energy conservation. Indeed the following identities
\begin{subequations}
\begin{align}
  \int \mathcal{Q}_{\alpha\beta}  \,m_\alpha v \, dv &= \nu_{\alpha\beta} m_\alpha n_\alpha \left( \overline{u}_\beta - u_\alpha \right) \,, \\
  \int \mathcal{Q}_{\alpha\beta}  \,m_\alpha \frac{v^2}{2} \, dv &= \nu_{\alpha\beta} \left( \frac{D_v}{2} n_\alpha k_B \left( \overline{T}_\beta - T_\alpha \right)  + \frac{1}{2}m_\alpha n_\alpha \left(\overline{u}^2_\beta- u_\alpha^2 \right) \right)\,
\end{align}
\end{subequations}
hold true for the operators defined by \eqref{eq:def:BGK}. The trivial choice $(\overline{u}_\beta,\overline{T}_\beta)=({u}_\beta,{T}_\beta)$ do ensure the plasma total momentum conservation,  provided that $\nu_{ei} m_e n_e = \nu_{ie} m_i n_i$. However in this case, the plasma total energy is not conserved. We refer to \cite{greene_improved_1973} for a seminal work, as well as \cite{klingenberg_consistent_2016} and the references therein for recent advances, on the choice of these parameters compliant with the desired properties \eqref{eq:prop:BGK:inter} of the inter-specie collision operators.

The electromagnetic field $(E,B)$ evolution is driven by the Maxwell system:
\begin{align}
& \frac{1}{c^2}\partial_t E - \nabla_x \times B= - \mu_0 J\,,\label{eq:std:Amp}\\
& \partial_t B+ \nabla_x \times E = 0\,,\label{eq:std:Far}\\
&\nabla_x \cdot E= \frac{\rho}{\epsilon_0}\,,\label{eq:std:divE}\\
& \nabla_x \cdot B = 0\,,\label{eq:std:divB}
\end{align}
where $c$ is the speed of light, $\mu_0$ the vacuum permeability and $\epsilon_0$ the vacuum permittivity verifying $\mu_0\epsilon_0 c^2 = 1$. The Maxwell sources are the particle currents and densities
\begin{subequations}
\begin{align}
  \rho &= q (n_i - n_e) \,, \\
  J &= q (n_i u_i - n_e u_e) \,.
\end{align}  
\end{subequations}

The definition of the collision frequencies as stated by Eqs.\eqref{eq:Freq:Ordering} relates different time scales. Indeed, because of their different masses, ions and electrons are not equally affected by collisions. This properties are more clearly emphasized working with dimensionless variables as proposed in the next section.

\subsection{Scaling of the Vlasov-Maxwell system}\label{sec:scaling}
\label{sec-scaling}

The equations are written with dimensionless quantities in order to easily identify different regimes.  The scaling is introduced under {\it a priori} assumptions that the electronic and ionic temperatures, densities and mean velocities are comparable with a magnitude denoted $T_0$, $n_0$ and $u_0$. These scales define the typical Debye length as well as the electron plasma period 
\begin{equation*}
\lambda_D=\sqrt{\frac{\epsilon_0 k_B T_0}{q^2 n_0}}\,, \qquad \tau_{pe}=\sqrt{\frac{m_e \epsilon_0}{q^2 n_0}}\,.
\end{equation*}
We denote by $x_0$ and $t_0$ the characteristic space and time scales of the phenomena observed, which yields to the velocity of interest $\vartheta_0 = x_0/t_0$. The magnitude of the thermal velocity for the specie $\alpha$ is denoted $v_{0,\alpha}$ with $v_{0,\alpha}^2=k_B T_0/m_\alpha$. Due to the different masses, the thermal veolicity of the electron is not that of the ions. The reference thermal velocity $v_0$ will be defined by the ion one  $v_{0}^2=k_B T_0/m_i$, hence $v_{0,e}= v_0/\varepsilon$ and $v_{0,i}=v_0$, where $\varepsilon^2=m_e/m_i$. Finally, the particle current scale is defined as $J_0 = q n_0 u_0$.
The dimensionless variables are defined according to 
\begin{gather*}
  x^* = \frac{x}{x_0} \,, \quad t^* = \frac{t}{t_0} \,,\quad v^* = \frac{v}{v_{0,\alpha}} \,, \quad f^* = \frac{f}{n_0/(v_{0,\alpha})^{D_v}} \,, \quad n^* = \frac{n}{n_0} \,,\quad J^* = \frac{J}{q n_0 u_0} \,, \\ E^* = \frac{E}{E_0} \,,\quad B^* = \frac{B}{B_0} \,, 
\end{gather*}
 the collision frequencies verifying
\begin{equation}\label{eq:scaling:relation:freq}
  \nu_{ee,0} = \nu_{ei,0} = \frac{1}{\varepsilon} \nu_{ii,0}\,, \qquad  \nu_{ie,0} = \varepsilon \nu_{ii,0}\,, \qquad \varepsilon = \sqrt{\frac{m_e}{m_i}}\,.
\end{equation}
On the fastest time scales, the electron distribution function relaxes towards a Maxwellian. On the same time scale, the electron mean velocity and temperature relax towards that of those of the ions. The relaxation of the ionic distribution function towards the local equilibrium is slower, by a factor $\varepsilon^{-1}=\sqrt{m_i/m_e}$. Finally, the ions are almost unaffected by the collisions against the electrons. The relaxation of the ionic distribution function towards that of the electrons define the largest time scale, by a factor $\varepsilon^{-1}$ compared to the the relaxation towards the thermodynamical equilibrium.

The dimensionless ionic and electronic Vlasov equations can be rewritten as (keeping the same notations for dimensionless variables): 
\begin{align}
  \begin{split}
&\xi \partial_t f_i +  v \cdot \nabla_x f_i + \eta (E+ \frac{\beta}{\xi} v \times B)\cdot \nabla_v f_i =\\
&\hspace*{6cm} \frac{ \xi}{\kappa} \Big( \nu_{ii} \left(\mathcal{M}_{n_i,u_i,T_i} - f_i\right) + \varepsilon \nu_{ie} \left( \overline{\mathcal{M}}_{n_i,u_e,T_e} - f_i \right) \Big) \,,\label{VMadim:a}\\  
  \end{split} \\ 
  \begin{split}
&\xi \varepsilon \partial_t f_e + v \cdot \nabla_x f_e - \eta(E+ \frac{\beta}{\varepsilon \xi} v \times B)\cdot \nabla_v f_e = \\
& \hspace*{6cm}  \frac{ \xi}{\kappa} \Big( \nu_{ee} \left(\mathcal{M}_{n_e,u_e,T_e} - f_e\right) + \nu_{ei} \left( \overline{\mathcal{M}}_{n_e,u_i,T_i} - f_e \right) \Big)\,,\label{VMadim:b}\\    
  \end{split}
\end{align}
together with the dimensionless Maxwell system writting
\begin{subequations}\label{eq:Maxwell:adim}
\begin{align}
& \lambda^2 \eta (\alpha^2\frac{\partial {E}}{\partial t} - \beta \nabla_x \times B)=-\alpha^2 \frac{M}{\xi} J\,,\label{eq:Maxwell:adim:Ampere}\\
& \beta \partial_t B + \nabla_x \times E = 0\,,\label{VMadim:c}\\
&\lambda^2 \eta \nabla_x \cdot E= n_i - n_e\,,\\
& \nabla_x \cdot B = 0\,,\label{VMadim:e}\\
& J = n_i u_i - n_e u_e \,.
\end{align}
\end{subequations}
This system is written thanks to the following dimensionless parameters
\begin{equation}
\left\{  \begin{array}[c]{l}
          \displaystyle \varepsilon^2=\frac{m_e}{m_i} \text{ the ratio of the electronic and ionic masses}\,,\\[3mm]
          \displaystyle  \lambda= \frac{\lambda_D}{x_0} \text{ the scaled Debye length}\,,\\[3mm]
          \displaystyle M = \frac{u_0}{v_{0}} \text{ the ionic Mach number, with } v_{0} = \sqrt{\frac{k_BT_0}{m_i}} \text{the ionic speed of sound}  \,,\\[3mm]
           \displaystyle \xi=\frac{\vartheta_0}{v_0} \text{ the ratio of the typical velocity to the ionic speed of sound}\,, \\[3mm]
         \displaystyle \alpha=\frac{\vartheta_0}{c} \text{ the ratio of the typical velocity to the speed of light}\,, \\[3mm]
         \displaystyle \eta=\frac{q x_0 E_0}{k_B T_0} \text{ the ratio of the electric and plasma internal energies} \,, \\[3mm]
         \displaystyle \beta=\frac{\vartheta_0B_0}{E_0} \text{ the induced electric field relative to the total electric field}\,,\\[3mm]        
           \kappa^{-1} = \nu_{ii,0} t_0 \text{ the number of ion-ion collisions during the typical time}\,.
  \end{array} \right.
\end{equation}

The dimensionless Maxwellians are defined by
\begin{subequations}\label{eq:def:Maxwellians}
\begin{align}
  \mathcal{M}_{n_e,u_\alpha,T_\alpha} = \overline{\mathcal{M}}_{n_e,u_\alpha,T_\alpha}  &= n_e(x,t) \left(\frac{1}{2\pi T_\alpha(x,t)}\right)^\frac{D_v}{2} \exp\left(-\frac{|M\varepsilon u_\alpha(x,t)-v |^2}{2T_\alpha(x,t)} \right) \,,\\
  \mathcal{M}_{n_i,u_\alpha,T_\alpha} = \overline{\mathcal{M}}_{n_i,u_\alpha,T_\alpha}  &= n_i(x,t) \left(\frac{1}{2\pi T_\alpha(x,t)}\right)^\frac{D_v}{2} \exp\left(-\frac{|M u_\alpha(x,t)-v |^2}{2T_\alpha(x,t)} \right) \,.
\end{align}  
\end{subequations}

Some comments can be stated regarding the meaning of these parameters and the scaling relations. 

The typical mean velocity and temperature are assumed to be the same for the electrons and the ions. Accordingly, the relaxation of the electron mean velocity and temperature towards that of the ions  may be assume to marginally contribute to the evolution of the system. This assumption is therefore consistent with the investigation of resistive-less plasma modellings and the neglect of the inter-species collisions.

The parameter $\xi$ is intended to provide a measure of how the electronic and ionic dynamics are resolved. The choice $\xi=1$ means that the system is assumed to evolve at a speed comparable to the ionic thermal velocity $v_0$, while $\xi\varepsilon=1$ performs a rescaling of this typical velocity to the electron microscopic velocity. Setting $\xi =M$ relates the typical speed of the system to the ionic mean velocity $u_0$. Actually, the Mach number measures the gap between the microscopic (thermal) and macroscopic velocity scales.

The scaling relation $\eta =1$ is generally assumed in single fluid plasma representation. The plasma internal energy is then on a par with the electric energy. This equilibrium is fundamental in the derivation of the Boltzmann relation. The identity $\beta M = \xi$ is also common in single-fluid plasma models. This amounts to assume that the induced electric field scales as the product of the plasma mean velocity and the typical magnetic field $E_0= u_0 B_0$. In other words, the magnetic field is essentially transported with the plasma flow. This later assumption is in line with the Alfven's frozen theorem \cite{moreau_magnetohydrodynamics_1990,freidberg_ideal_2014,davidson_introduction_2001,schnack_lectures_2009} characteristic of ideal MHD models: the magnetic field is frozen into the plasma and transported by its flow. 

The derivation of reduced models consists in identifying small dimensionless parameters and let them go to zero. The smallness of the scaled Debye length refers to a typical space scale much larger than the physical Debye length. This means that it assumed that the charge separations occurring on space scales comparable to the Debye length are not important to explain the evolution of the system. Sending the scaled Debye length to zero performs a low frequency filtering into the equation deriving thus a quasi-neutral model.
In the context of the derivation of numerical methods, the typical length relates to the mesh size. This outlines the advantage of reduced models: the low frequency filtering operated by vanishing small parameters permits to derive numerical methods with discretization parameters (mesh size and time step) unconstrained by the small scales filtered out from the original equations.

\section{A hierarchy of quasi-neutral models bridging the Vlasov-Maxwell system and the Hall-MHD regime}\label{sec:fluid:hierarchy}

\subsection{Handling the fluid and quasi-neutral limits}

\subsubsection{Introduction}
\begin{figure}[!ht]
  \centering
  \includegraphics[width=0.7\textwidth]{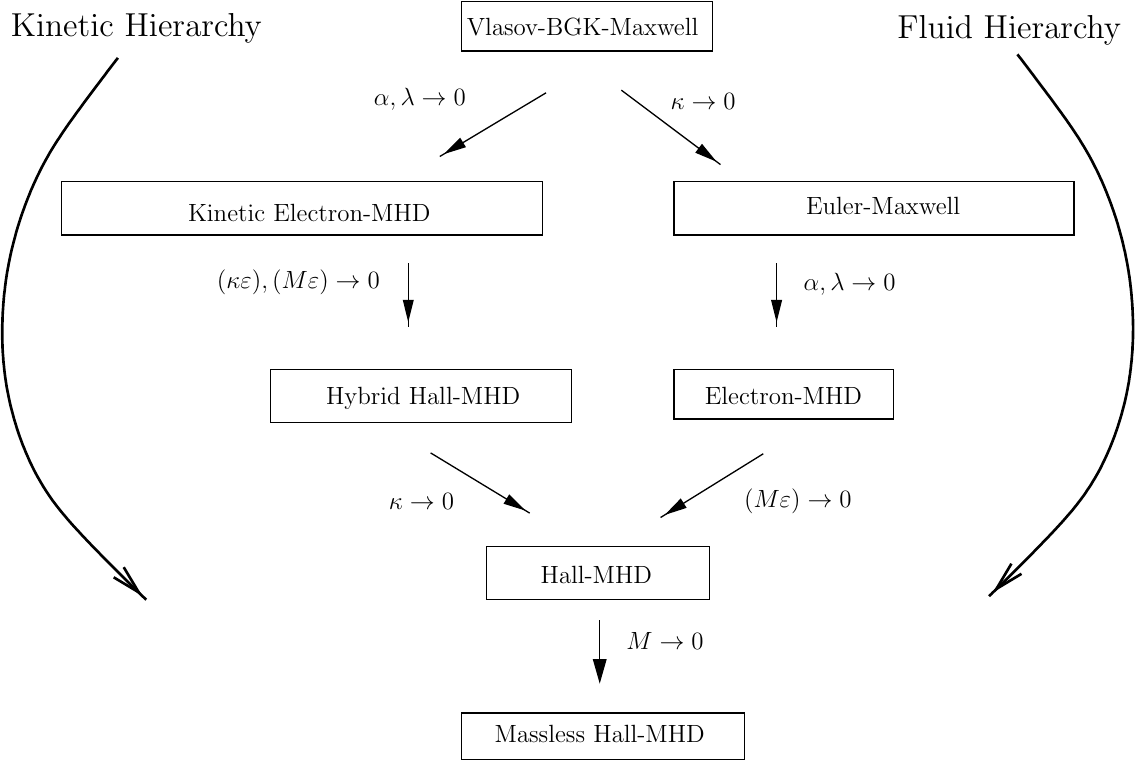}
  \caption{Fluid and kinetic (quasi-neutral) model hierarchies derived from the Vlasov-BGK-Maxwell system.}
  \label{fig:hierarchy}
\end{figure}

The aim here is to reduce the number of free dimensionless parameters deriving by this means different reduced models well suited for the description of low frequency phenomena. 
As depicted in Fig.~\ref{fig:hierarchy}, the starting point of this hierarchy of models implements the minimal upgrades of the Vlasov-Maxwell system to recover a MHD regime. Precisely the only inter species collisions are taken into account in the initial model in order for the distribution function to relax towards the local equilibrium. This yields
\begin{subequations}\label{eq:sys:Boltzmann}
\begin{align}
  \begin{split}
&\xi \partial_t f_i +  v \cdot \nabla_x f_i + \eta (E+ \frac{\beta}{\xi} v \times B)\cdot \nabla_v f_i =\frac{ \xi}{\kappa}\nu_{ii} \left(\mathcal{M}_{n_i,u_i,T_i} - f_i\right)  \,,\label{VMadimBis:a}\\  
  \end{split} \\ 
  \begin{split}
&\xi \varepsilon \partial_t f_e + v \cdot \nabla_x f_e - \eta(E+ \frac{\beta}{\varepsilon \xi} v \times B)\cdot \nabla_v f_e =\frac{ \xi}{\kappa}  \nu_{ee} \left(\mathcal{M}_{n_e,u_e,T_e} - f_e\right) \,,\label{VMadimBis:b}\\    
  \end{split}
\end{align}
\end{subequations}
for the evolution of the ions and electrons coupled to the dimensionless Maxwell system defined by Eqs.~\eqref{eq:Maxwell:adim}.

From the scaling relations stated by Eq.~\eqref{eq:scaling:relation:freq}, discarding the inter-species collisions make sense for the ions. Due to their large mass, the ions are almost unaffected by encounters with electrons. 
For the electrons, this assumption is not in line with the scaling of the like and inter-species collision frequencies. However, the purpose here, is to propose a physically meaningful framework to clarify the foundation of a numerical method bridging the gap between a kinetic description of a weakly (or non) collisional magnetized plasma with a MHD regime. The interspecies collisions give rise to the resistivity in the macroscopic system which is not the targeted class of modelling for this work.

\subsubsection{Handling the fluid limit}
To identify easily a fluid regime, the distribution function is decomposed into a Maxwellian $\mathcal{M}_{n_\alpha,u_\alpha,T_\alpha} $ and a deviation from this Maxwellian ${\kappa} \, g_\alpha$ according to
  \begin{subequations}\label{def:micro-macro}
\begin{equation}
  f_\alpha = \mathcal{M}_{n_\alpha,u_\alpha,T_\alpha} + {\kappa} \, g_\alpha \,,
\end{equation}
the deviation verifying 
\begin{equation}
 \big< v^n g_\alpha \big> = \int_{\Omega_v} v^n g_\alpha \, dv =0\,, \quad n=0,1,2\,.
\end{equation}    
  \end{subequations}
With this decomposition, the Vlasov-Boltzmann equations \eqref{eq:sys:Boltzmann} can be recast into a hydrodynamic set of equations with kinetic corrections, depending on the moment of the deviation $g_\alpha$, yielding
\begin{subequations}\label{eq:Sys:MM:ions}
\begin{align}
  &\frac{\xi}{M} \partial_t n_i + \nabla_x \cdot (n_i u_i) = 0 \label{eq:cons:ni}\\
  \begin{split}
  &M^2\left(\frac{\xi}{M} \partial_t (n_i u_i) + \nabla_x \cdot (n_iu_i \otimes u_i) \right)+ \nabla_x p_i - \eta \, n_i \big( E + \frac{\beta M}{\xi} u_i \times B \big) =\\
  & \hspace*{11cm}- {\kappa}\nabla_x \cdot \left< v \otimes v g_i \right> \,,    
  \end{split} \\
& \frac{\xi}{M}  \partial_t W_i + \nabla_x \cdot \big((W_i + p_i)u_i \big) - \eta n_i E\cdot u_i = -\frac{\kappa}{M} \nabla_x \cdot \left< \frac{|v|^2}{2} v g_i\right>\,.
\end{align}
with 
\begin{equation}
  W_i = (M)^2 \frac{1}{2} n_i u_i^2 + \frac{p_i}{\gamma-1} \,, \qquad p_i = n_i T_i \,,
\end{equation}
\end{subequations}
for the ions, and an equivalent system for the electrons,
\begin{subequations}\label{eq:Sys:MM:electrons}
\begin{align}
  &\frac{\xi}{M} \partial_t n_e + \nabla_x \cdot (n_e u_e) = 0 \label{eq:cons:ne}\\
  \begin{split}
  &\left(M\varepsilon\right)^2\left(\frac{\xi}{M} \partial_t (n_e u_e) + \nabla_x \cdot (n_eu_e \otimes u_e) \right)+ \nabla_x p_e + \eta\, n_e \big(  E + \frac{\beta M}{\xi}u_e \times B \big) =\\
  & \hspace*{11cm}- {\kappa}\nabla_x \cdot \left< v \otimes v g_e \right> \,,    
  \end{split} \\
& \frac{\xi}{M}  \partial_t W_e + \nabla_x \cdot \big((W_e + p_e)u_e \big) + \eta n_e E\cdot u_e = -\frac{\kappa}{M\varepsilon} \nabla_x \cdot \left< \frac{|v|^2}{2} v g_e\right>\,.
\end{align}
with 
\begin{equation}
  W_e = (M\varepsilon)^2 \frac{1}{2} n_e u_e^2 + \frac{p_e}{\gamma-1}\,, \quad p_e = n_e T_e\,.
\end{equation}
\end{subequations}
These two systems are coupled to a set of equations (the Maxwell system \eqref{eq:Maxwell:adim}) driving the changes in the electromagnetic field.

\subsubsection{On the quasi-neutral limit}
Omitting the collisions, the fastest velocity in this system is the propagation of waves at the speed of light described by the Maxwell system. The Debye length as well as the plasma period also  define small space and time scales for large plasma densities.
The quasi-neutral limit is defined by the following scaling relations:
\begin{equation}\label{eq:def:QN:limit}
 (\alpha,\lambda) \to 0\,, \quad  {\alpha}\sim{\lambda} \,.
\end{equation}
This amounts to assume that the scaled Debye length is small compared to the typical length and that the system evolves at a speed lower than the speed of light. By this means, the small scales related to these parameters are filtered out of the equations. The last hypothesis $\alpha\sim \lambda$ is essential to recover the low frequency Ampere's law, derived by neglecting the displacement current. This equation being common to Magneto-Hydro-Dynamic models, the quasi-neutral limit encompasses this two assumptions. With the vanishing of this generalised dimensionless Debye length $(\lambda,\alpha)\to 0$, the Maxwell system degenerates into
\begin{subequations}\label{eq:Maxwell:Deg}
\begin{align}
& \beta \nabla_x \times B = \frac{M}{\xi} J\,,\label{eq:Maxwell:Deg:Ampere}\\
& \beta \partial_t B + \nabla_x \times E = 0\,,\label{eq:Maxwell:Deg:Faraday}\\
& n_i = n_e\,,\\
& \nabla_x \cdot B = 0\,.\label{VMadim:e}
\end{align}
\end{subequations}
From Gauss's law, the property of the electronic density to match that of the ions is recovered, which genuinely enforces the quasi-neutrality of the plasma. The electric field has no contribution in both these degenerate Gauss and the Ampere equations. The remaining occurrence of the electric field is limited to the Faraday equation \eqref{eq:Maxwell:Deg:Faraday}. Therefore, 
this set of equations is not well suited for the computation of the electric field. Indeed, the electrostatic component of the electric field can be arbitrarily chosen in Eqs.~\eqref{eq:Maxwell:Deg}. 

In the quasi-neutral limit, the electric field is provided by the particle current $J$ rather than displacement current ($\partial E/\partial t$ originally present in Ampere's law). To close the system, the dependence of $J$ with respect to $E$ shall be explained to restore uniqueness of the electric field. This is related to the model describing the plasma.

\subsection{A hierarchy of kinetic models for quasi-neutral plasmas}

\subsubsection{A kinetic formulation of the Electron-MHD}\label{sec:kinetic:EMHD}
The aim here is to follow the microscopic dynamic of the electrons. The velocity of interest is the kinetic velocity of the electrons. This amounts to set $\vartheta_0 = v_0/\varepsilon$ or equivalently $\xi \varepsilon = 1$, yielding
\begin{subequations}\label{VMadimEMHD}
\begin{align}
  \begin{split}
&\partial_t f_i + \varepsilon \left(v \cdot \nabla_x f_i + \eta (E+ {\varepsilon\beta} v \times B)\cdot \nabla_v f_i\right) =\frac{1}{\kappa}\nu_{ii} \left(\mathcal{M}_{n_i,u_i,T_i} - f_i\right)  \,,\label{VMadimEMHD:a}\\  
  \end{split} \\ 
  \begin{split}
& \partial_t f_e + v \cdot \nabla_x f_e - \eta(E+ {\beta} v \times B)\cdot \nabla_v f_e =\frac{1}{\varepsilon \kappa}  \nu_{ee} \left(\mathcal{M}_{n_e,u_e,T_e} - f_e\right) \,,\label{VMadimEMHD:b}\\    
  \end{split}
  \end{align}
  \end{subequations}
The collisions are assumed to be ineffective on the characteristic time scale:  $\kappa\varepsilon\gg{1}$ which amounts to neglect the collision operator in Eqs.~\eqref{VMadimEMHD}, in particular for the ions, owing to $\varepsilon \ll 1$. 

At these scales, the ions may be considered at rest.
Performing the quasi-neutral limit $(\lambda=\alpha)\to 0$, the system at hand here is recast into (see \cite{DegDelDoy})
\begin{align}
&\partial_t f_e + v \cdot \nabla_x f_e -\eta (E+\beta v \times B)\cdot \nabla_v f_e =0\,,\label{QNVM1}\\
& \beta \nabla_x \times B=(M\varepsilon)J\,,\label{QNVM2}\\
& \beta\partial_t B + \nabla_x \times E = 0\,,\label{QNVM3}\\
&n_e= n_i=n\,,\label{QNVM4}\\
& \nabla_x \cdot B = 0\,.\label{QNVM5}
\end{align}
First, note that the formal time derivative of the Faraday equation \eqref{QNVM3} together with the curl of Ampere's law yields
  \begin{equation}\label{eq:ampere:farraday:0}
    \nabla_x \times \nabla_x \times E  = - (M \varepsilon) \partial_t J \,,
  \end{equation}
which outlines that the electric field is known up to the gradient of a potential in this system. 
In \cite{DDSprep,DegDelDoy} the ill-posed nature of this equation is corrected by explaining the relation between the current density and the electric field. The first moment of Eq.~\eqref{QNVM1} yields the conservation of the electronic momentum, with
\begin{subequations}
\begin{equation}\label{momentJ}
   (M \varepsilon) \frac{\partial}{\partial t} (n u_e)  = - \nabla_x \cdot \mathbb{S} - \eta \big( n E +  \beta (M \varepsilon) n u_e \times B  \big)=0\,, \qquad 
\end{equation}
with $n=n_e=n_i$ and
\begin{equation}
  \mathbb{S}= \int \left(v \otimes v \right)\, f_e \, dv \,.
\end{equation}
\end{subequations}
Inserting this identity into Eq.~{\eqref{eq:ampere:farraday:0} gives 
\begin{equation}\label{eq:ampere:QN}
   \eta n  E + \nabla_x \times \nabla_x \times E = \eta \beta (M\varepsilon)J\times B - \nabla_x \cdot \mathbb{S} .
\end{equation}
This equation is well posed in the quasi-neutral limit ($n>0$) and can be used for the computation of the electric field.  This yields the following definition of the quasi-neutral model
\begin{subequations}\label{eq:def:sys:EMHD}
\begin{align}
&\partial_t f_e + v \cdot \nabla_x f_e -(E+ v \times B)\cdot \nabla_v f_e =0\,,\label{QN1bis}\\
& n E  + \nabla_x \times \nabla_x \times E = (M\varepsilon)J\times B - \nabla_x \cdot \mathbb{S}\,,\label{QN2bis}\\\
&\partial_t B + \nabla_x \times E = 0\,,\label{QN3bis}\\
& \nabla_x \cdot B = 0\,.\label{QN5bis}
\end{align}
\end{subequations}
This model is written under the assumption $\eta=1$ meaning that the plasma thermal energy is on a par with the electric energy together with  $\beta=1$.

Note that the electric field provided by Eq.~\eqref{QN2bis} enforces a divergence free particle current, or more precisely $\partial_t (\nabla \cdot J) = 0$. This yields, thanks to the continuity equation:
\begin{equation*}
\frac{\partial^2 \rho}{\partial t^2} = 0 \,.
\end{equation*} 
This proves the consistency of this model with the quasi-neutrality assumption (matching of the electronic and ionic densities) as soon as the initial data are compliant with this regime. The evolution of the ions may be taken into account thanks to another Vlasov equation. This brings corrections in deriving Eq.~\eqref{QN2bis} accounting for the contribution  of the ionic particle current.

The characteristics of this models are similar to the so-called Electron MHD: the time scale of interest is that of the electrons, the ions merely creating a motionless background for the fast electron flows \cite{kingsep_electron_1990}.  In particular, this modelling accounts for the inertia of electrons. A noticeable difference with the Electron-MHD (see Sec.~\ref{Sec:EMHD}) lies in the kinetic description of the plasma. An asymptotic-Preserving method is proposed in \cite{DegDelDoy} to bridge this quasi-neutral model and the Vlasov-Maxwell system. The properties of this quasi-neutral plasma description are investigated in \cite{tronci_neutral_2015} by means of a linear stability analysis.

\subsubsection{A hybrid formulation of the Hall-MHD}\label{sec:hybrid}
Hybrid modelling \cite{buchner_hybrid_2003,yin_hybrid_2002,tronci_hybrid_2014} refers to a class of plasma models where the ions are described by a kinetic equation while the fluid limit is assumed for the electrons. This is in line with the scaling relations of the collision frequencies stated by Eqs.~\eqref{eq:scaling:relation:freq}. The relaxation of the electronic distribution function towards the local equilibrium is indeed faster than for the ions. The aim of these modelling is to filter out of the equations the fast scales carried by the electron dynamics. Therefore, a zero inertia regime is also assumed for the electrons together with the fluid limit and the quasi-neutrality of the plasma.

The typical velocity selected here is the microscopic (thermal) velocity of the ions. This translates into the identity $\xi = 1$ resulting in the following system for the plasma:
\begin{subequations}\label{VMadim:Hybrid}
\begin{align}
  \begin{split}
&\partial_t f_i + v \cdot \nabla_x f_i + \eta (E+ \frac{\beta}{\xi} v \times B)\cdot \nabla_v f_i =\frac{1}{\kappa}\nu_{ii} \left(\mathcal{M}_{n_i,u_i,T_i} - f_i\right)  \,,\label{VMadim:Hybrid:a}\\  
  \end{split} \\ 
  \begin{split}
& \partial_t f_e + \frac{1}{\varepsilon}\left( v \cdot \nabla_x f_e - \eta(E+ \frac{\beta}{\varepsilon \xi} v \times B)\cdot \nabla_v f_e \right)=\frac{1}{\varepsilon \kappa}  \nu_{ee} \left(\mathcal{M}_{n_e,u_e,T_e} - f_e\right) \,,\label{VMadim:Hybrid:b}\\    
  \end{split}
  \end{align}
  \end{subequations} 
The fluid limit for the electrons is selected assuming $(\varepsilon \kappa) \ll 1$ meaning that the number of electron collisions during the typical time is large. The quasi-neutrality of the plasma amounts to set $\lambda=\alpha\ll1$.
To overcome the degeneracy of the Maxwell system in the quasi-neutral limit, the electronic momentum is harnessed to provide the so-called generalised Ohm's law. The electronic system can be recast into

\begin{align*}
\begin{split}
 &\left(M\varepsilon\right)^2\left(\frac{1}{M} \partial_t (n_e u_e) + \nabla_x \cdot (n_eu_e \otimes u_e) \right)+ \nabla_x p_e + \eta\,\Big( n_e  E + (\beta M) u_e\times B\Big) = \\
  & \hspace*{9.5cm }- (\kappa (M\varepsilon)) \nabla_x \cdot \sigma_e \,,         
  \end{split}\\
   \begin{split}
& \frac{1}{M}  \partial_t W_e + \nabla_x \cdot \Big((W_e + p_e)u_e \Big) + \eta n_e E\cdot u_e = 
-\frac{\kappa}{M\varepsilon}\nabla_x \cdot \Big( \left(M\varepsilon \right)^2\sigma_e \cdot u_e + \mu_e \nabla_x T_e \Big)\,;
\end{split} 
\end{align*}
with, owing to the quasi-neutrality assumption, $n_e = n_i = n$.

The dynamic described by these equations is stiff, this is due to the smallness of $(\xi\varepsilon)$ in this regime: the thermal velocity of the ions (defined as the typical velocity) is small compared to that of the ions. Therefore, the electrons are in a low Mach regime. Assuming $(M\varepsilon)\ll 1$ gives rise to the following equilibria
\begin{subequations}\label{sys:forces:equilibria}
\begin{align}
 &\nabla_x (n T_e) + \eta\,n \Big(  E + (\beta M) u_e\times B\Big) = 0 \,,  \\
 & \nabla_x T_e = 0        \,.
 \end{align}
\end{subequations}

The classical massless approximation for the electrons is recovered with the generalised Ohm's law and a homogeneous electronic temperature. The definition of the mean velocity $u_e$ is derived from the particle current density  $ J= n (u_i - u_e)$ together with Amp\`ere's law \eqref{QNVM2}, yielding 
\begin{equation}\label{eq:subs:ue:ui:j}
  u_e = u_i - \frac{\beta}{M}\frac{\nabla_x \times B}{n} \,,
\end{equation}

The hybrid plasma modelling writes (assuming $\eta=\beta=1$)
\begin{subequations}
\begin{align}
&\partial_t f_i + v \cdot \nabla_x f_i + (E+  v \times B)\cdot \nabla_v f_i =0  \,,\\  
& E = - M u \times B+ M \frac{\nabla_x \times B}{n}\times B - T_e \frac{\nabla_x n}{n}\,,\\
&\frac{\partial B}{\partial t} + \nabla_x \times E = 0 \,, \qquad \nabla_x \cdot B = 0 \,,
\end{align}
with
\begin{align}
  n = \int f_i \, dv \,, \quad n u = \int v f_i \, dv \,.
\end{align}
\end{subequations}
The derivation of a similar model is proposed in \cite{acheritogaray_kinetic_2011} with numerical investigations in \cite{degond_simulation_2011}. 

\subsection{A fluid hierarchy of quasi-neutral models}

\subsection{The Electron MHD-system}\label{Sec:EMHD}
This model is obtain by letting $\kappa \to 0$ in Eqs.~\eqref{eq:sys:Boltzmann}. This yields the following set of equations for the electrons
\begin{subequations}\label{eq:Sys:Bifluide:electrons}
\begin{align}
  &\frac{\xi}{M} \partial_t n_e + \nabla_x \cdot (n_e u_e) = 0 \label{eq:bifluid:ne}\\
  \begin{split}
  &\left(\frac{\xi}{M} \partial_t (n_e u_e) + \nabla_x \cdot (n_eu_e \otimes u_e) \right)+ \frac{1}{(M\varepsilon)^2} \left( \nabla_x p_e + \eta\, n_e (  E + \frac{\beta M}{\xi}u_e \times B ) \right) =0\,,    
  \end{split} \\
& \frac{\xi}{M}  \partial_t W_e + \nabla_x \cdot \big((W_e + p_e)u_e \big) + \eta n_e E\cdot u_e = 0\,.
\end{align}
\end{subequations}
A similar system  is derived for the ions however with $\varepsilon=1$ and $\eta$ replaced by $-\eta$. These two sets of conservation laws are coupled to the Maxwell system \eqref{eq:Maxwell:adim}. Performing the quasi-neutral limit in this system $(\lambda=\alpha)\to 0$ and focusing on the electronic dynamics with $\xi\varepsilon=1$ yields the quasi-neutral bi-fluid Euler-Maxwell system also referred to as Electron-MHD system. 
This model is similar to that of Sec.~\ref{sec:kinetic:EMHD} but with a fluid description for the plasma. It is implemented and investigated numerically in the framework of Asymptotic-Preserving methods in \cite{DDSprep}.

\subsubsection{The Hall-MHD regime}\label{sec:Hall:MHD}
The Hall-MHD regime (see \cite{lighthill_studies_1960,witalis_hall_1986,schnack_lectures_2009}) is recovered from the assumptions of the precedent section but with a typical velocity equal to the plasma mean flow yielding $\xi=M$. The fast electronic dynamics is filtered out from the equations to provide a low frequency modelling for the plasma driven by the evolution of the massive ions. The plasma velocity, denoted $u$ is defined as that of the heavy species $u=u_i$.
The other parameters obey the classical scaling relations of MHD models: $\eta=1$ and $\beta=1$. 

In the drift regime ($M\varepsilon\to0)$, the electronic energy reduces to the internal energy 
\begin{subequations}
  \begin{equation}
    \mathcal{E}_e= p_e/(\gamma-1)\,,
  \end{equation}
 with the electronic momentum and energy verifying
  \begin{align}
    &\nabla_x p_e = - n \left(E + u_e \times B \right)\,,\label{eq:qe:QN}\\
    &\partial_t \mathcal{E}_e + \nabla_x \cdot \big( (\mathcal{E}_e+p_e) u_e \big) = - n E\cdot u_e \,.\label{eq:We:QN}
  \end{align}
  \end{subequations}
The generalysed Ohm's law \eqref{eq:qe:QN}  is harnessed to compute the electric field. The electronic velocity $u_e$ is substituted by $u_e=u - J/n$.

The electric field is computed thanks to the generalized Ohm's law, giving rise to
\begin{equation}\label{eq:Ohm:law}
  E = - u \times B+ \frac{\nabla_x \times B}{n}\times B - \frac{\nabla_x p_e}{n}\,.
\end{equation}
The first term of the right hand side of the equation \eqref{eq:Ohm:law} is the classical frozen field term, explaining the convection of the magnetic field together with the plasma . The second and third contributions are the so-called Hall and diamagnetic terms.
Inserting this definition in the Faraday law \eqref{eq:Maxwell:Deg:Faraday}, the magnetic induction equation can be constructed, with
\begin{equation}\label{eq:magnetic:induction:Hall}
  \partial_t B + \nabla_x \cdot (u \otimes B - B \otimes u) = - \nabla_x \times \left(\frac{\nabla_x \times B}{n}\times B  \right) + \nabla_x \times\left( \frac{\nabla_xp_e}{n}\right)\,.
\end{equation}

Finally the plasma mass density, momentum and total pressure $p$ and energy $W$ defined by 
\begin{subequations}
\begin{align}
  p &= p_i +p_e = n (T_i+T_e)  \,,\\
  W &= W_i + \mathcal{E}_e = (M)^2 \frac{1}{2}  n u^2 + \frac{p}{\gamma-1} \,,
\end{align}
verify
\begin{align}
  &\partial_t n + \nabla_x \cdot (n u) = 0 \\
  & \partial_t (n u) + \nabla_x \cdot (nu \otimes u) + \frac{1}{M^2} \nabla_x p = \frac{1}{M^2} J \times B \\
  & \partial_t W + \nabla_x \cdot \big( (W + p) u \big) + \nabla_x \cdot \Big( (\mathcal{E}_e+p_e) \vHall \Big) = E\cdot J \,,
\end{align}
\end{subequations}
the Hall velocity $\vHall$, which can be interpreted as the electron velocity in the ion frame, is defined by
\begin{equation}
  \vHall = - \frac{J}{n}\,.
\end{equation}
The ideal MHD equations are classically written under a conservative form using the system total pressure and energy
\begin{subequations}
  \begin{equation}
  p_\text{TOT} = p + \frac{B^2}{2} \,,\qquad   W_\text{TOT} = W + \frac{B^2}{2}  \,,
  \end{equation}
writing the system as 
\begin{align}
   &\partial_t n + \nabla_x \cdot (n u) = 0 \\
  & \partial_t (n u) + \nabla_x \cdot \Big( nu \otimes u - \frac{1}{M^2} B \otimes B \Big) + \frac{1}{M^2} \nabla_x p_\text{TOT} = 0 \\
  & \partial_t W_\text{TOT} + \nabla_x \cdot \Big( (W_\text{TOT} + p_\text{TOT}) u - (B\cdot u) B \Big) =  - \nabla_x \cdot \Big( (\mathcal{E}_e+p_e) \vHall \Big)\,,\\
  & \partial_t B - \nabla_x \cdot \Big( B\otimes (u+\vHall) - (u+\vHall) \otimes B \Big) = \nabla_x \times \left( \frac{\nabla_xp_e}{n}\right)\,.
\end{align}
This set of equations is supplemented with the electronic energy conservation \eqref{eq:We:QN}. 
\end{subequations}

The ideal MHD equations are recovered from this system assuming an ideal Ohm's law where the current density is assumed small compared to the ion mean velocity and therefore neglected. However, in this simplified framework (omitting the unlike particle collisions), there are no mechanisms preventing the electron mean velocity to depart from that of the ions. Consequently, the generalized Ohm's law incorporates the Hall velocity in complement to the so called ideal Ohm's law. The effect of the resistivity should be consider to derive the ideal mHD regime.

The drift approximation operated for the electrons amounts to vanishing the electronic Mach numbers $(M\varepsilon)$. The scale separation introduced by the small electron to ion mass ratio $\varepsilon$ is not always sufficient to consider this limit independently to vanishing ionic Mach numbers $M \to 0$. For low ionic Mach number a low frequency filtering may be operated performing the limit of vanishing electronic Mach numbers jointly with the ionic Mach numbers. This asymptotic defined the Massless MHD regime \cite{besse_model_2004}.

\section{The electrostatic regime and the Boltzmann relation}\label{sec:electrostatic}

\subsection{Electrostatic limit of the Maxwell system}

The electrostatic regime is recovered from the dimensionless Maxwell system \eqref{eq:Maxwell:adim} by letting $\alpha$ go to zero. This assumption shall be interpreted as a typical velocity negligible compared to the speed of light. From Amp\`ere's law
\begin{equation}\label{eq:Ampere:electro}
    \lambda^2 \eta \frac{\partial {E}}{\partial t} + \frac{M}{\xi} J =\frac{\beta  \lambda^2 \eta}{\alpha^2} \nabla_x \times B \,,
\end{equation}
the limit $\alpha\to 0$ provides $\nabla \times B=0$. Nonetheless, the right hand side of Eq.~\eqref{eq:Ampere:electro} remains an undetermined form. Therefore the Ampere's law is not well suited for the computation of the electric field in the electrostatic limit.
However, subjected to convenient boundary conditions, the property $\nabla_x \times B = 0$ together with $\nabla_x \cdot B = 0$ yields $\partial_t B = 0$. Inserting this property into the Faraday equation \eqref{VMadim:c} provides an electrostatic electric field: $\nabla \times E = 0$. Furthermore, the undetermined form in Eq.~\eqref{eq:Ampere:electro} is divergence free. Therefore, computing the divergence of Ampere's law provides
\begin{equation*}
  \lambda^2 \eta \frac{\partial \nabla_x \cdot {E}}{\partial t} = -  \frac{M}{\xi} \nabla_x \cdot J \,,
\end{equation*}
which is a well posed problem for the electric field under the condition $\nabla_x \times E = 0$. Note that, owing to the continuity equation
\begin{equation*}
  \frac{\partial \rho}{\partial t} + \frac{M}{\xi} \nabla_x \cdot J = 0\,,
\end{equation*}
originating from the conservation of the particle densities \eqref{eq:cons:ni} and \eqref{eq:cons:ne}, the divergence of 
Amp\`ere's law is equivalent to the time derivative of Gauss law, with
  \begin{equation*}
    \frac{\partial}{\partial t} \left(\lambda^2 \eta \nabla_x \cdot E -{\rho}\right)=0\,.
  \end{equation*}
Therefore, in the electrostatic regime, the Gauss equation is used to compute the electric field.

\subsection{Quasi-neutral models at the electronic scale}
This analysis is carried out under the assumption of a vanishing magnetic field ($B=0$). The plasma description consider in the sequel is therefore
\begin{subequations}
\begin{align}
  &\xi \partial_t f_i +  v \cdot \nabla_x f_i + \eta E \cdot \nabla_v f_i = \frac{ \xi}{\kappa} \Big( \nu_{ii} \left(\mathcal{M}_{n_i,u_i,T_i} - f_i\right) \Big) \,,\\  
  & \xi \varepsilon \partial_t f_e + v \cdot \nabla_x f_e - \eta E\cdot \nabla_v f_e = \frac{ \xi}{\kappa} \Big( \nu_{ee} \left(\mathcal{M}_{n_e,u_e,T_e} - f_e\right) \Big)\,\label{eq:VB:electrostatic}
\end{align}
 coupled to the Gauss equation 
\begin{equation}
- \lambda^2 \eta \Delta \phi= n_i - n_e\,,
\end{equation}
$\phi$ being the electrostatic potential, with $E = - \nabla \phi$ and $n_\alpha = \int f_\alpha dv$. 
\end{subequations}
The quasi-neutrality of the plasma is recovered for vanishing scaled Debye length $\lambda \to 0$. In this regime, similarly to the electromagnetic case, an equation needs to be manufactured from the motion of the particles, to compute the electric field. This is classically obtained thanks to the equation of the electronic momentum conservation. The electric field is computed in order for this conservation to be satisfied. 
In \cite{DDNSV10} an equivalent approach is proposed. It consists in using the time derivatives of Gauss law to produce
\begin{equation}\label{eq:Gauss:double:derivative}
- \lambda^2 \eta \frac{\partial^2}{\partial t^2} \Delta \phi = \frac{\partial^2 }{\partial t^2} \left( n_i -n_e\right) \,,
\end{equation}
In the quasi-neutral limit ($\lambda\to 0$), the electrostatic field is the Lagrangian multiplier of the quasi-neutrality constraint
\begin{equation*}
\frac{\partial^2 }{\partial t^2} \left( n_i -n_e\right) = 0\,.
\end{equation*}
From the system~\eqref{eq:Sys:MM:electrons}, the following identity is recovered
\begin{align}
&\left(\frac{\xi}{M}\right)^2\frac{\partial^2 n_e}{\partial t^2} = -\nabla_x \cdot \Big(\nabla_x \cdot (n_e u_e \otimes u_e) \Big)- \frac{1}{(M\varepsilon)^2} \nabla_x \cdot \Big(\nabla_x P_e + \eta n_e E\Big) \,.\label{eq:dt2:ne} \\
&P_e = p_e \Id + \kappa \left< v \otimes v g_e \right> \,, \qquad \left< v \otimes v \varphi \right> = \int \varphi \, dv \,. 
\end{align}
Resuming the scaling relations of Sec.~\ref{sec:kinetic:EMHD}: $\xi=1/\varepsilon$ and $\kappa \varepsilon > 1$, assuming that the ions are at rest, the evolution of the charge density $n_i -n_e$ is governed by Eq.~\eqref{eq:dt2:ne} with
\begin{equation}\label{eq:dt2:ne:electrostatic}
\frac{\partial^2 n_e}{\partial t^2} = -(M\varepsilon)^2 \nabla_x \cdot \Big(\nabla_x \cdot (n_e u_e \otimes u_e) \Big)- \nabla_x \cdot \Big(\nabla_x P_e + \eta n_e E\Big) \,.\\
\end{equation}
The evolution of the density is barely independent of the mean flow velocity but relies on the balance between the pressure and the electric forces. The equation providing the electric potential $\phi$ is obtained by inserting this relation into Eq.~\eqref{eq:Gauss:double:derivative} and passing to the limit $\lambda \to 0$.

This yields the following quasi-neutral kinetic plasma description
\begin{subequations}
\begin{align}
 &\partial_t f_e + v \cdot \nabla_x f_e + \eta \nabla_x \phi \cdot \nabla_v f_e = \frac{1}{\varepsilon\kappa} \Big( \nu_{ee} \left(\mathcal{M}_{n_e,u_e,T_e} - f_e\right)\,,\label{eq:electrostatic:kinetic}\\
&- \eta \nabla_x \cdot \left( n_e \nabla_x \phi \right) = - \nabla_x \cdot \Big( \nabla_x \cdot P_e - (M\varepsilon)^2 \nabla_x \cdot (n_e u_e \otimes u_e) \Big) \,.
\end{align}
According the values of $\kappa$, the collision term in Eq.~\eqref{eq:electrostatic:kinetic} may be disregarded, defining therefore a non collisional kinetic description. Contrariwise, letting $\kappa \to 0$, a fluid description for the electrons may be derived, with 
\end{subequations}

\begin{subequations}\label{eq:sys:electrostatic:electron:fluid}
\begin{align}
  &\frac{1}{(M\varepsilon)} \partial_t n_e + \nabla_x \cdot (n_e u_e) = 0 \label{eq:sys:electrostatic:electron:density}\\
  \begin{split}
  &\frac{1}{(M\varepsilon)} \partial_t (n_e u_e) + \nabla_x \cdot (n_eu_e \otimes u_e) + \frac{1}{(M\varepsilon)^2} \left( \nabla_x p_e -\, \eta n_e \nabla_x \phi  \right) =0\,,  \label{eq:sys:electrostatic:electron:momentum}  
  \end{split} \\
& \frac{1}{(M\varepsilon)}  \partial_t W_e + \nabla_x \cdot \big((W_e + p_e)u_e \big) - \eta n_e \nabla_x\phi\cdot u_e = 0\,.
\end{align}
\end{subequations}
Note that a similar equation to \eqref{eq:dt2:ne:electrostatic}, but with $P_e = p_e \Id$, may be worked out of the conservation of the electronic density \eqref{eq:sys:electrostatic:electron:density} and momentum \eqref{eq:sys:electrostatic:electron:momentum}. This outlines that the electronic dynamics, in particular the electronic speed of sound, is resolved in this model. Comparable models are implemented  and numerical experiences in the context of the Asymptotic-Preserving methods in \cite{CDV07} for fluid plasma description and \cite{manfredi_vlasov_2011} for kinetic equations.
\subsection{Quasi-neutral models at the ionic scale}
The typical velocity is chosen to be that of the ions with either $\xi=1$ for the kinetic descriptions of the ions and $\xi=M$ for macroscopic models.

The hybrid modelling investigated in Sec.~\ref{sec:hybrid} is defined by the scaling relations $\xi=1$, $(\varepsilon \kappa)\ll 1$ and $\eta =1$. The equilibria stated by Eqs.~\eqref{sys:forces:equilibria} yields:
\begin{equation}
  T_e \nabla_x n_e = n_e \nabla_x \phi \,,
\end{equation}
with an homogeneous electronic temperature. This equation is integrated to provide the so-called Boltzmann relation
\begin{equation}\label{eq:Boltzmann:relation}
  n_e = n_0 \exp\left( - \frac{\phi}{T_e}\right) \,,
\end{equation}
$n_0$ being a constant (independent of the space variable $x$) that should be determined from adequate conditions \cite{hagelaar_how_2007}. Due to the Boltzmann relation, the quasi-neutral limit is not singular any more. Indeed plugging the Boltzmann relation \eqref{eq:Boltzmann:relation} into the Gauss equation yields
\begin{equation}
- \lambda^2 \Delta \phi= n_i - n_0 \exp\left(-\frac{\phi}{T_e}\right)\,.
\end{equation}
This equation is not degenerate for the computation of the electric potential for vanishing $\lambda$. Indeed, the non linear part of the equation provides a means of computing $\phi$ in the quasi-neutral limit. This property is thoroughly investigated  in \cite{degond_numerical_2012}.

The hybrid electrostatic model may be recast into
\begin{subequations}\label{eq:sys:Valsov:Boltzmann}
\begin{align}
&\partial_t f_i +  v \cdot \nabla_x f_i - \nabla_x \phi \cdot \nabla_v f_i =  \frac{1}{\kappa} \Big( \nu_{ii} \left(\mathcal{M}_{n_i,u_i,T_i} - f_i\right) \Big)\,,\\
& \phi = - T_e \ln\left(\frac{n_i}{n_0}\right) \,.
\end{align}
\end{subequations}

Letting $\kappa \to 0$ together with $\kappa/(M\varepsilon)\ll 1$ and $\xi=M$ yields the quasi-neutral fluid model 
\begin{subequations}\label{eq:sys:Euler:Boltzmann}
\begin{align}
  &\partial_t n_i + \nabla_x \cdot (n_i u_i) = 0 \\
  &\partial_t (n_i u_i) + \nabla_x \cdot (n_iu_i \otimes u_i) + \frac{1}{M^2} \left(\nabla_x p_i + n_i \nabla_x\phi\right)  =0 \,,    \\
&  \partial_t W_i + \nabla_x \cdot \big((W_i + p_i)u_i \big) + n_i \nabla_x\phi\cdot u_i = 0\,,\\
& \phi = - T_e \ln\left(\frac{n_i}{n_0}\right) \,.
\end{align}
\end{subequations}
In the models \eqref{eq:sys:Valsov:Boltzmann} and \eqref{eq:sys:Euler:Boltzmann} following the evolution of the plasma at the ionic scale, the fast electronic dynamics introduced by the electron inertia is filtered out of the equations by performing the low frequency limit $(M\varepsilon) \to 0$.
\section{Conclusions}\label{sec:conclusions}

In this paper, we propose an asymptotic analysis bridging kinetic plasma descriptions coupled to the Maxwell system and single plasma modelling. Two frameworks are investigated. The first one is devoted to electromagnetic fields. The plasma is represented by a hierarchy of models starting with the bi-kinetic Vlasov-Maxwell system while ending with the single fluid Hall-Magneto-Hydro-Dynamic model. The second framework is dedicated to electrostatic fields. In this context, the asymptotic analysis permits to derive a hierarchy of models bridging the bi-kinetic Vlasov-Poisson system to a single fluid representation consisting of a fluid system for the ions coupled to the Boltzmann relation for the electrons.
The investigations proposed within this document unravel different asymptotic parameters explaining the transition from one model to the other. The effort conducted in the present work consists in relating these asymptotic parameters to characteristics of the system. This means that the transition from one model to the other may be explained by a change in the plasma characteristics or the typical scales at which the plasma is observed. 

This last notion is important in the perspective of designing a numerical method. Indeed, the discretization of these equations requires the use of a mesh interval as well as a time step. These two numerical parameters define the typical space and time scale, therefore a velocity scale as well, the numerical method is aimed at capturing. This is related to the parameter $\xi$ used for the asymptotic analysis. Regarding the quasi-neutral modellings investigated within this document, different choices are operated for this parameter. The fastest scales are related to the electron thermal velocity when the fast electron dynamics is intended to be captured by the model. This is for instance the value selected for Electron-MHD regimes, either in the fluid or kinetic frameworks. For hybrid or single fluid plasma representations, the velocity scale is reduced to that of the mean flow of the plasma defined by the massive ions. The organisation of Sec.~\ref{sec:electrostatic} is aimed at emphasizing this feature.

The second parameter, already established in precedent work (see \cite{DegDelDoy,DDSprep,DDNSV10}), is the generalized scaled Debye length $\lambda$. It actually encompasses the scaled Debye length and the ratio of the typical velocity to the speed of light. Vanishing the generalised Debye length amounts to filter out from the equations the small scales attached to the charge separation as well as those related to the propagation of electromagnetic waves at the speed of light. The quasi-neutral limit is therefore a low frequency limit. Quasi-neutrality breakdowns may be explained by a refinement of the typical length scale or, for instance, a decrease of the plasma density. This changes are well accounted for by the asymptotic parameters set up to perform the analysis.

The vanishing of the electron inertia is related to a low Mach regime $(M\varepsilon)\ll 1$. 
In single fluid plasma representation, the fast electron dynamic is dropped out of the equations to perform a low frequency filtering, the system being assumed to evolve at a lower speed attached to the massive ions. Nonetheless, the nature of the flow may be subjected to significant changes explaining that the particle inertia becomes significant again to account for the system evolution. This is illustrated in studies of plasma flows in sheaths, with supersonic particles, while the mean plasma velocity is small compared to the speed of sound in the plasma bulk \cite{chodura_plasma_1986,stangeby_plasma_2000,GMAH08,DGAH10,manfredi_plasma-wall_2008}. Accounting for this phenomena is possible selecting the appropriate typical velocity to resolve or filter the fast electron dynamic.

Finally the fluid assumption is classically related to a vanishing of the Knudsen number $\kappa\ll 1$ accounting for the relaxation of the distribution function towards the local thermodynamic equilibrium. The interplay of these four dimensionless parameters ($\xi$, $\lambda$, $M\varepsilon$, $\kappa$) define a hierarchy of reduced models bridging kinetic plasma descriptions coupled to the Maxwell system to quasi-neutral plasma representations including kinetic, hybrid and single fluid modellings. These later models, namely the Hall-MHD and the Boltzmann relation are widely used to design efficient numerical methods. The asymptotic analysis conducted within this document draws the guidelines for the derivation of numerical methods implementing local up-scalings, therefore widening the use of numerical methods discretizing these reduced models.

 \section*{Acknowledgments} 
This work has been supported by ``F\'ed\'eration de Fusion pour la Recherche par Confinement Magn\'etique'' (FrFCM) in the frame of the project 
"BRIDIPIC: BRIDging Particle-In-Cell methods and low frequency numerical models of plasmas".\newline
A. Crestetto acknowledges support from the french ``Agence Nationale pour la Recherche (ANR)'' in the frame of the projects 
MoHyCon ANR-17-CE40-0027-01 and MUFFIN  ANR-19-CE46-0004.
\appendix

\section{Micro-Macro decomposition, computation of the viscous terms.}
\subsection{Introduction, definitions, elementary properties}
The analysis carried out in this section are developed in the electrostatic framework and specified for the electrons. The extensions for either electromagnetic fields or the ions are straightforward and are therefore omitted for conciseness.

We first introduce the projector onto the Maxwellian
For any smooth function $\varphi$, the projector onto the Maxwellian denoted $\Me$ with
\begin{subequations}\label{eq:def:PiMe:Me}
  \begin{equation}\label{eq:def:Me}
    \Me = n_e(x,t) \frac{1}{\big(2\pi T_e(x,t)\big)^{D_v/2}} \exp\left(\frac{\left(v - (M \varepsilon) u_e(x,t) \right)^2}{2 T_e(x,t)} \right) \,.
  \end{equation}
 For any smooth function $\varphi$, the projector onto $\Me$, denoted $\PiMe$, is defined as (see \cite{bennoune_uniformly_2008,crestetto_kinetic/fluid_2012})
\begin{equation}
  \begin{split}    
  &\PiMe\left(\varphi \right) = \Bigg[\left< \varphi \right> + \frac{\big(v - (M\varepsilon)u_e\big)}{T_e}\cdot \left< \big(v - (M\varepsilon)u_e \big) \varphi \right> \\
   & \hspace*{2.5cm}+ \frac{2}{D_v}\left(\frac{\left| v - (M\varepsilon)u_e\right|^2}{2 T_e}- \frac{D_v}{2}\right) \left<  \left(\frac{\left| v - (M\varepsilon)u_e\right|^2}{2 T_e}- \frac{D_v}{2}\right) \varphi \right>\Bigg]\frac{\mathcal{M}_e}{n_e}\,.
\end{split}
\end{equation}  
where $\left< \varphi \right> = \int \varphi \, dv$.
\end{subequations}
For $k= 1,\ldots,D_v$, we have the following properties
\begin{subequations}\label{lemma:ge:utils}
\begin{alignat}{3}
  &(\Id - \PiMe)(\partial_t\Me) = (\Id - \PiMe)(E\cdot \nabla_v\Me) =0 \,,\\
  &\left(\Id - \PiMe \right) \left( \frac{1}{n_e} \Me \, v_k \partial_{x_k} n_e \right)  = 0 \,;\\
  \begin{split}
   &   \left(\Id - \PiMe \right) \left( \big(v - (M\varepsilon) u_e \big) \cdot \partial_{x_k}u_e \, v_k \Me \right) = \Big(- \frac{\big(v - (M\varepsilon) u_{e}\big)^2 }{D_v}\partial_{x_k}u_{e,k} \\
   & \hspace*{6.5cm}+ \big(v_k - (M\varepsilon) u_{e,k}\big) \big(v - (M\varepsilon) u_k\big) \cdot \partial_{x_k}u_e
 \Big) \Me \,;     
  \end{split}\\
  \begin{split}
    & \left(\Id - \PiMe \right) \left( \Big(\frac{(v - (M\varepsilon)u_e)^2}{2 T_e^2} - \frac{D_v}{2 T_e} \Big) \Me \, v_k \partial_{x_k} T_e  \right) =\\
    & \hspace*{4cm}\Me \left( \frac{\big(v - (M\varepsilon)u_e\big)^2}{2 T_e^2} - \frac{D_v+2}{2 T_e}\right) \Big( v_k - (M\varepsilon) u_{e,k} \Big)   {\partial_{x_k} T_e}\,.
  \end{split}
\end{alignat}
Furthermore, if $g_e$ satisfies the Micro-Macro decomposition \eqref{def:micro-macro}, the following identities holds true
\begin{align}
  &\PiMe(g_e) = \PiMe(\partial_t g_e) = 0 
\end{align}  
\end{subequations}

\subsection{Computation of the deviation to the Maxwellian}

The aim here is to characterize $g_e$ or more specifically an approximation to the first order in $\kappa$.
Inserting the micro-macro decomposition into the Vlasov-Boltzmann equation \eqref{eq:VB:electrostatic} yields
 \begin{equation*}
   \xi \varepsilon \partial_t \Me + v \cdot \nabla_x \Me - \eta E\cdot \nabla_v \Me + \kappa \left( \xi \varepsilon \partial_t g_e + v \cdot \nabla_x g_e - \eta E\cdot \nabla_v g_e  \right) = \LMe g_e\,,
\end{equation*}
where
\begin{equation*}
  \LMe g_e = -\xi \nu_{ee} g_e \,.
\end{equation*}
This provides, using properties~\eqref{lemma:ge:utils},
\begin{equation*}
  \LMe g_e =  (\Id - \PiMe)\left( v\cdot \nabla_x \Me \right) +\kappa \Big(\xi \varepsilon \partial_t g_e+ (\Id - \PiMe)\big( v\cdot \nabla_x g_e - \eta E \cdot \nabla_v g_e \big) \Big)\,.
\end{equation*}
It follows
\begin{subequations}
\begin{equation}
     g_e = \left( \LMe\right) ^{-1} \Big( (\Id - \PiMe)\left( v\cdot \nabla_x \Me \right) + \mathcal{O}(\kappa)  \Big)\,,
\end{equation} 
with
\begin{equation}
  v_k \partial_{x_k} \Me = \Bigg(\frac{ \partial_{x_k} n_e}{n_e} + \frac{M \varepsilon }{T_e} \Big(v - (M\varepsilon) u_e \Big) \cdot \partial_{x_k}u_e +
 \Big(\frac{(v - (M\varepsilon)u_e)^2}{2 T_e^2} - \frac{D_v}{2T_e} \Big) \partial_{x_k} T_e \Bigg) v_k \Me \,.
\end{equation}  
\end{subequations}
From properties~\eqref{lemma:ge:utils}, we can state the expression of the deviation to the Maxwellian
\begin{equation}\label{eq:def:ge}
  \begin{split}
    g_e &= -\frac{\Me}{\xi \nu_{ee}} \sum_{k=1}^{D_v}\Bigg(\frac{M \varepsilon}{T_e}\Big(- \frac{\big(v - (M\varepsilon) u_{e}\big)^2 }{D_v}\partial_{x_k}u_{e,k} \\
    & \hspace*{1.5cm} + \big(v_k - (M\varepsilon) u_{e,k}\big) \big(v - (M\varepsilon) u_e\big) \cdot \partial_{x_k}u_e  \Big)  \\
    & \hspace*{2.5cm}+ \left( \frac{\big(v - (M\varepsilon)u_e\big)^2}{2 T_e^2} - \frac{D_v+2}{2 T_e}\right) \Big( v_k - (M\varepsilon) u_{e,k} \Big)   {\partial_{x_k} T_e} \Bigg)+ \mathcal{O}\left(\frac{\kappa}{\xi} \right)\,.
\end{split}
\end{equation}

\subsection{Computation of the viscous terms}

The viscous terms are defined by
\begin{align*}
  \left< v\otimes v g_e \right> &= - \frac{M\varepsilon}{\xi} \sigma_e +  \mathcal{O}\left(\frac{\kappa}{\xi} \right)\,, \\
  \left< \frac{|v|^2}{2}v g_e \right> &= - \frac{1}{\xi} \Big( (M \varepsilon)^2 \sigma_e\cdot u_e  + q_e \Big) + \mathcal{O}\left(\frac{\kappa}{\xi} \right) \,.
\end{align*}
Following the characterization \eqref{eq:def:ge} of $g_e$ we can write

  \begin{align*}
    \begin{split}
    \sigma_e &=  - \frac{1}{T_e}\frac{1}{\nu_{ee}} \Bigg( \left< -\frac{\big(v - (M\varepsilon) u_{e}\big)^2 }{D_v}  \Me \big( v\otimes v \big)\right> \sum_{l=1}^{D_v} \partial_{x_l}u_{e,l} \\      
 &\hspace*{2cm} + \sum_{l=1}^{D_v} \Big<(v\otimes v)\Me\big(v_l - (M\varepsilon) u_{e,l}\big)\sum_{k=1}^{D_v} \big(v_k - (M\varepsilon) u_{e,k}\big) \partial_{x_l}u_{e,k} \Big> \Bigg)\,,
    \end{split}\\
    q_e & = - \frac{1}{\nu_{ee}}\left< \frac{|v|^2}{2} v \Me \left( \frac{\big(v - (M\varepsilon)u_e\big)^2}{2 T_e^2} - \frac{D_v+2}{2 T_e}\right) \Big( v - (M\varepsilon) u_{e} \Big)   \cdot {\nabla_x T_e}   \right>\,,    
  \end{align*}
Inserting in these definitions, the following identities 
\begin{align*}
  & \left< \frac{|v|^2}{2} (v\otimes v) \Me \left( \frac{\big(v - (M\varepsilon)u_e\big)^2}{2 T_e^2} + \frac{D_v+2}{2 T_e}\right) \Big( v - (M\varepsilon) u_{e} \Big)   \cdot {\nabla_x T_e}   \right> = 0 \,, \\
  \begin{split}
  &  -\frac{1}{T_e}\frac{1}{\nu_{ee}}\Bigg( \left< \frac{\big(v - (M\varepsilon) u_{e}\big)^2 }{D_v}  \big( v\otimes v \big) \Me\right> \sum_{l=1}^{D_v} \partial_{x_l}u_{e,l} \\      
 &\hspace*{0.5cm} - \sum_{l=1}^{D_v} \Big<(v\otimes v)\Me\big(v_l - (M\varepsilon) u_{e,l}\big)\sum_{k=1}^{D_v} \big(v_k - (M\varepsilon) u_{e,k}\big) \partial_{x_l}u_{e,k} \Big>\Bigg) = (M\varepsilon) \sigma_e \cdot u_e \,;
  \end{split}
\end{align*}
we obtain
\begin{subequations}\label{lemma:q:sigma}
\begin{align}
  \sigma_e &= -\frac{1}{\nu_{ee}} \left( n_e T_e \right) \big( \nabla_x u_e + \nabla_x u_e ^T - \frac{2}{D_v}\left(\nabla_x \cdot u_e\right) \Id \big) \,,\\
  q_e &= -\frac{D_v+2}{2 \nu_{ee}} \left( n_e T_e\right) \nabla_x T_e\,.
\end{align}

\end{subequations}

\bibliographystyle{abbrv}
\bibliography{plasma}


\end{document}